\documentstyle[12pt]{article}
\let\LARGE=\large

\newcommand{\alt}{\mathrel{\raisebox{-.6ex}{$\stackrel{\textstyle<}{\sim}$}}}
\newcommand{\agt}{\mathrel{\raisebox{-.6ex}{$\stackrel{\textstyle>}{\sim}$}}}
\def\overlay#1#2{\ifmmode \setbox 0=\hbox {$#1$}\setbox 1=\hbox to\wd 0{\hss
$#2$\hss }\else \setbox 0=\hbox {#1}\setbox 1=\hbox to\wd 0{\hss #2\hss }\fi
#1\hskip -\wd 0\box 1}
\def\case#1/#2{{\textstyle{#1\over#2}}}
\def\detected{{\rm detected}}
\def\Dirac{{\rm [Dirac]}}
\def\etal{{\it et al}}
\def\L{{\rm L}}
\def\Majorana{{\rm [Majorana]}}
\def\sec{{\rm s}}
\def\total{{\rm tot}}
\def\visible{{\rm vis}}
\catcode`@=11
\newcount\@tempcntc
\def\@citex[#1]#2{\if@filesw\immediate\write\@auxout{\string\citation{#2}}\fi
  \@tempcnta\z@\@tempcntb\m@ne\def\@citea{}\@cite{\@for\@citeb:=#2\do
    {\@ifundefined
      {b@\@citeb}{\@citeo\@tempcntb\m@ne\@citea\def\@citea{,}{\bf ?}\@warning
      {Citation `\@citeb' on page \thepage \space undefined}}%
    {\setbox\z@\hbox{\global\@tempcntc0\csname b@\@citeb\endcsname\relax}%
     \ifnum\@tempcntc=\z@ \@citeo\@tempcntb\m@ne
       \@citea\def\@citea{,}\hbox{\csname b@\@citeb\endcsname}%
     \else
      \advance\@tempcntb\@ne
      \ifnum\@tempcntb=\@tempcntc
      \else\advance\@tempcntb\m@ne\@citeo
      \@tempcnta\@tempcntc\@tempcntb\@tempcntc\fi\fi}}\@citeo}{#1}}
\def\@citeo{\ifnum\@tempcnta>\@tempcntb\else\@citea\def\@citea{,}%
 \ifnum\@tempcnta=\@tempcntb\the\@tempcnta\else
  {\advance\@tempcnta\@ne\ifnum\@tempcnta=\@tempcntb \else \def\@citea{--}\fi
   \advance\@tempcnta\m@ne\the\@tempcnta\@citea\the\@tempcntb}\fi\fi}
\catcode`@=12

\font\fortssbx=cmssbx10 scaled \magstep2
\hbox to \hsize{
\hskip.5in \raise.1in\hbox{\fortssbx University of Wisconsin - Madison}
\hskip2in$\vcenter{\hbox{\bf MADPH-95-875}
                   \hbox{\bf RAL-95-026}
                   \hbox{\bf OUTP-95-09P}
                   \hbox{\bf hep-ph/9503295}
                   \hbox{April 1995}}$ }
\vspace{.75in}
\begin{document}
\thispagestyle{empty}
\begin{center}
{\huge Remarks on the KARMEN anomaly}\\[.4in]
V.~Barger$^a$, R.J.N.~Phillips$^b$ and S.~Sarkar$^c$\\[.2in]
\it
$^a$Physics Department, University of Wisconsin, Madison, WI 53706, USA\\
$^b$Rutherford Appleton Laboratory, Chilton, Didcot, Oxon OX11 0QX, UK\\
$^c$Theoretical Physics, University of Oxford, 1 Keble Road, Oxford OX1 3NP, UK
\end{center}
\vspace{.5in}
\begin{abstract}
A recently reported anomaly in the time structure of signals in the
KARMEN neutrino detector suggests the decay of a new particle $x$,
produced in $\pi^+ \to \mu^+ x$ with mass $m_x=33.9$ MeV. We discuss
the constraints and difficulties in interpreting $x$ as a neutrino.
We show that a mainly-sterile neutrino scenario is compatible with all
laboratory constraints, within narrow limits on the mixing parameters,
although there are problems with astrophysical and cosmological
constraints. This scenario predicts that appreciable numbers of other
$x$-decay events with different origins and time structures should
also be observable in the KARMEN detector.  Such $x$-decay events
should also be found in the LSND experiment and may be relevant to the
search for $\bar\nu_\mu\to\bar\nu_e$ oscillations.
\end{abstract}
\vspace{.75in}
\begin{center}
(Published in {\sl Phys. Lett. B352 (1995) 365}; erratum added June 1995)
\end{center}
\newpage
The KARMEN collaboration, which studies the interactions of neutrinos
from the stopped $\pi^+$ decay chain at RAL, has recently reported an
anomaly~\cite{kar} in the time-dependence of their signals.  This
anomaly suggests the production of a new weakly-interacting neutral
particle (call it $x$) in the initial $\pi^+$ decays, which travels
with well determined velocity $\beta_x = v_x/c \simeq 1/60$ and decays
in the detector after a mean flight path of 17.5 m.  The distinctive
feature of the $x$-events is their timing, apparently at a well
determined interval $3.6 \mu$s after the arrival and prompt decay of
the pion pulse (which determines $\beta_x$); however, the visible
energy in the detector scintillator shows no anomaly, so $x$ decays
apparently deposit visible energy similar to typical neutrino
interactions. The present note briefly discusses the interpretation of
$x$ as a massive neutrino. We show that a mainly-sterile neutrino
scenario is compatible with all laboratory constraints, for either
Dirac or Majorana options, within rather narrow bounds on the mixing
parameters. There are some problems with astrophysical and cosmological
constraints, but it is interesting nevertheless to explore the
further implications for laboratory experiments which can test
this interpretation directly. This scenario predicts that appreciable
numbers of other $x$-decay events with different origins and time
structures should also be observable in the KARMEN detector. Such
$x$-decay events should also be found in the LSND experiment at
LAMPF~\cite{lsnd} and may be relevant to the ongoing search for
$\bar\nu_\mu\to\bar\nu_e$ oscillations~\cite{osc}.

If we postulate no other new particles below the pion mass, then the
precise time structure~\cite{kar} requires $x$ production to go via
one of the two-body modes $\pi^+\to\mu^+x$ or $\pi^+\to e^+x$ .
However the latter implies mass $m_x=137.2$ MeV (determined from
$\beta_x$) and hence anomalously large visible $x$-decay energy, with
mean value $<T_{\visible}>\; \simeq 51$ MeV for $x\to\mu e \nu \to
ee\nu\nu\nu$ or $<T_{\visible}>\; \simeq 88$ MeV for $x\to ee\nu$,
compared to neutrino interactions that typically give $T_{\visible} \sim
11-35$ MeV~\cite{kar}. We must therefore presume that $x$ is produced
via $\pi^+\to \mu^+x$, with $m_x=33.9$ MeV determined from $\beta_x$.
Assuming no new weak interactions, the standard decay mode is $x\to
e^-e^+\nu_e$; note that flavour-changing neutral current (FCNC)
processes $x\to\nu\nu\nu$ and $x\to\nu\gamma$ are highly suppressed
for a standard isodoublet neutrino (we consider the isosinglet case
later)~\cite{deruj,hprs}.

The neutrino charged-current eigenstates $\nu_{\alpha {\L}}$
($\alpha =e,\mu,\tau$) which appear in the weak interaction
coupled via $W$ to $e,\mu,\tau$, may be written as coherent
superpositions of mass eigenstates $\nu_{i {\L}}$ ($i=1,2,\ldots x$)
using the usual mixing matrix $U_{\alpha i}$ ;
\begin{equation}
\nu_{\alpha {\L}} = \Sigma_i U_{\alpha i} \nu_{i {\L}}\ .
\end{equation}
(In general $U$ is a $n\times n$ matrix, with $\alpha$ running over
all $SU(2)_L$ multiplet assignments and $i$ running over all masses).
Then the $x$ production and decay processes are scaled by factors
$|U_{\mu x}|^2$ and $|U_{e x}|^2$, respectively:
\begin{eqnarray}
\frac{\Gamma (\pi\to\mu x)}{\Gamma (\pi\to\mu\nu)}& = &
\frac{|U_{\mu x}|^2[m_\pi^2(m_\mu^2+m_x^2)-(m_\mu^2-m_x^2)^2]
      \lambda^{1\over 2}(m_\pi^2, m_\mu^2, m_x^2)}
     {m_\mu^2(m_\pi^2 - m_\mu^2)^2},\\
\frac{\Gamma (x\to e^+e^-\nu_e)}{\Gamma (\mu\to\nu_\mu\bar\nu_e e)}&=&
\frac{|U_{e x}|^2 m_x^5}{m_\mu^5}\;\ {\Dirac}\ , \quad
\frac{2|U_{e x}|^2 m_x^5}{m_\mu^5}\;\ {\Majorana}\ ,
\end{eqnarray}
where $\lambda (a,b,c) =a^2+b^2+c^2-2ab-2bc-2ca$ and we neglect
$m_e^2/m_x^2$; we recall that Majorana neutrinos decay twice as fast
as Dirac neutrinos with the same coupling, because their right chiral
components are not inert.  Hence the production branching fraction and
mean decay lifetime are simply given by
\begin{equation}
B(\pi\to\mu x) = 0.0285\, |U_{\mu x}|^2,
\label{eq:bf}
\end{equation}
\begin{equation}
\tau (x\to ee\nu) = 645\,|U_{ex}|^{-2} \; \mu {\sec}\;\ {\Dirac}\ ,
\quad
\tau (x\to ee\nu) = 323\,|U_{ex}|^{-2} \; \mu {\sec}\; {\Majorana}\ .
\label{eq:t}
\end{equation}
The correlation between branching fraction and lifetime, needed to
explain the KARMEN anomaly, was shown as a curve in the ($\tau ,B$)
plane in Ref.~\cite{kar}. We reproduce this as the solid curve in
Fig.1, extrapolating along the dashed curve with fixed $B/\tau$, and
showing also the scales of $|U_{ex}|^2$ and $|U_{\mu x}|^2$ implied by
Eqs.(\ref{eq:bf})-(\ref{eq:t}) along the upper and right-hand edges of
the diagram (the Dirac option is illustrated for $|U_{ex}|^2$).  The
regions with $|U_{\alpha x}|^2 > 1$ have no physical meaning in our
scenario.

Direct experimental constraints on the mixing elements $|U_{\alpha
i}|$ are summarized in Ref.~\cite{pdg}; they generally depend on mass
and the constraints we quote below are all for $m_x=33.9$ MeV.
Absence of a correction to the $\rho$ parameter of the $e$ spectrum in
$\mu\to e\nu\nu$ decay gives~\cite{shrock}
\begin{equation}
|U_{ex}|^2 \; + \; |U_{\mu x}|^2 \; < \; 2 \times 10^{-3}.
\label{eq:shrock}
\end{equation}
Absence of decay events in neutrino beams gives~\cite{berg,coop,bern}
\begin{equation}
|U_{ex}| \; |U_{\mu x}| \; < \; 1.5 \times 10^{-5}.
\label{eq:bb}
\end{equation}
Absence of anomalous contributions to $\pi\to e\nu$ gives~\cite{delee}
\begin{equation}
|U_{ex}|^2 \; < \; 0.85 \times 10^{-6}.
\label{eq:delee}
\end{equation}
Limits from neutrinoless $\beta\beta$-decay searches~\cite{maj} on the
effective $\nu_e$ Majorana mass\\ $\langle m_{\nu e} \rangle =
|\Sigma_j \eta_j m_j U_{ej}^2|$, where $\eta_j=\pm$ is the CP
signature of Majorana neutrino $\nu_j$, would require (see also
Ref.~\cite{moh})
\begin{equation}
|U_{ex}|^2 \; \alt \; 6 \times 10^{-8} \quad {\Majorana}\ ,
\label{eq:maj}
\end{equation}
where we have conservatively taken $\langle m_{\nu e} \rangle < 2$
eV~\cite{maj}. This bound would apply if $x$ were a Majorana state and
there were no substantial cancellations in the sum but not if $x$ were
a Dirac neutrino or part of a quasi-Dirac pair (with opposite CP
signatures). Other direct laboratory constraints are weaker than
these~\cite{pdg}. Studies of short muon tracks in $\pi\to\mu\to e$
events, from pions stopping in emulsion, would give stringent
constraints on $|U_{\mu x}|$ for $m_x < 33$ MeV~\cite{shrock}; but for
the present value $m_x=33.9$ MeV, the muon kinetic energy is only 1.5
keV giving an unobservable track length less than 1 micron, so no
constraint can be derived on this basis.  The $x$ mass and mixing
predict a contribution to the $\mu\to e\gamma$ branching
fraction~\cite{mueg}
\begin{equation}
B(\mu\to e\gamma) = 3\alpha m_x^2/(32\pi M_W^2) |U_{ex}^*U_{\mu x}|^2
                  \simeq \; 2.5\; \times \; 10^{-23},
\label{eq:mueg}
\end{equation}
a factor $\sim 2\times 10^{12}$ below the experimental upper
limit~\cite{pdg}. The constraints of
Eqs.(\ref{eq:shrock})-(\ref{eq:delee}) are shown on Fig.1; they leave
a range of ``solutions" to the KARMEN anomaly, based on $x\to e^-e^+\nu$
decay, described by
\begin{eqnarray}
|U_{ex}| \; |U_{\mu x}|  \; & \simeq & \; 0.8 \times 10^{-6}
\; {\Dirac}\ , \quad 0.6 \times 10^{-6} \; {\Majorana}\ ,\\
|U_{\mu x}| \; & \alt & 4.5 \times 10^{-2},\\
|U_{ex}|    \; & \alt & 10^{-3}\quad {\Dirac}\ , \quad
            2.5  \times 10^{-4}\quad {\Majorana}\ .
\label{eq:sol}
\end{eqnarray}
There are also constraints on the mass and identification of $x$.  The
ARGUS bound $m(\nu_\tau) < 31$ MeV~\cite{argus}, the CLEO bound
$m(\nu_\tau) < 32.6$ MeV~\cite{cleo} and the recent ALEPH bound
$m(\nu_\tau) < 24$ MeV~\cite{aleph} (all at $95\%$~C.L.), exclude $x$
from being the major component of $\nu_\tau$. Since LEP experiments
measure $N_\nu = 2.988\pm 0.023$~\cite{lep} light neutrino species
(weighted by their isodoublet mixing factors), the chiral component
$x_{\L}$ participating in standard weak interactions must then be
dominantly isosinglet, i.e. sterile.

An isodoublet interpretation of $x$ is also excluded by cosmological
and astrophysical arguments concerning unstable neutrinos in the mass
and lifetime range of interest~\cite{gelmini}. If $x$ has standard
weak interactions, its cosmological relic abundance would be
sufficiently high that its decay products would have distorted the
spectrum of the 2.73 K blackbody radiation background unless its
lifetime were less than $\sim 10^5$~s~\cite{dicus,sarkar}. This
requires $|U_{ex}|^2 \agt 10^{-8}$ (for $m_x=33.9$ MeV) and removes
part of the solution range in Eq.(\ref{eq:sol}). At such early times
the background photons are energetic enough to be Compton scattered by
the $e^+ e^-$ pairs from $x$ decay to energies above the $^2$H
photofission threshold and may thus undo
nucleosynthesis~\cite{sarkar}. Taking into account the energy
degradation due to $\gamma-\gamma$ scattering~\cite{lindley} this sets
a lifetime bound of $\alt 2 \times 10^{3}$~s~\cite{sarkar2}
corresponding to $|U_{ex}|^2 \agt 5\times 10^{-7}$, which leaves only
a tiny region of the remaining solution range. (Further constraints on
isodoublets from consideration of the entropy production by the
decaying particle are rather sensitive to the adopted upper limit to
the primordial $^4$He abundance~\cite{kolb}.) This loophole is closed
by consideration of the production and decays of massive neutrinos in
Supernova 1987A. For example the process $x \to \nu_e e^+ e^- \gamma$
operates at a rate $\alpha/2 \pi$ relative to the decay $x \to \nu_e
e^+ e^-$ and would have generated a $\gamma$ ray burst which was not
observed by the SMM satellite~\cite{dar}. When combined with other
arguments relating to energy deposition inside the
supernova~\cite{falk}, as well as experimental bounds on fast
$\nu_{\tau}$ decays~\cite{berg,coop}, this rules out {\it all} lifetimes
shorter than $\sim 10^8$~s~\cite{turner}, forbidding the entire
solution range for doublet neutrinos.

In order to evade the above bounds, it may appear adequate to require
$x$ to be mainly sterile since its direct production, both in
supernovae and in the early universe, is then suppressed. However,
sterile neutrinos can also be produced through their mixing with
doublet neutrinos, modulated by matter effects. During the collapse
phase of a supernova, resonant $\nu_e \to x$ conversions may cause
rapid deleptonisation of the core which would probably prevent the
supernova explosion ~\cite{raffelt}; however this does not happen for
$\Delta m^2 \agt 10^8$ eV$^2$ so will not apply to the $x$
particle. Secondly the energy loss due to emission of sterile
neutrinos during the cooling phase would have excessively shortened
the $\bar{\nu_e}$ burst from SN~1987A for a mixing in the range
$|U_{ex}|^2 \sim 10^{-9} - 10^{-2}$~\cite{kmp}; this argument applies
to a neutrino with mass up to $\sim 50-100$ MeV~\cite{raffelt2} (see
also Ref.~\cite{moh}) and will rule out the allowed region for
$x$. Most crucially, the absence of a $\gamma$ ray burst from SN 1987A
would, as in the case of doublet neutrinos discussed above, rule out
radiative decays~\cite{raffelt2} with lifetimes in the range $\sim
10^1-10^8$~s, taking into account the enhancement of radiative decays
for a singlet neutrino (see below). The production of sterile
neutrinos through neutrino oscillations in the early
universe~\cite{manohar} also presents a problem since $x$ particles
can be brought into thermal equilibrium at temperatures exceeding a
few GeV. Although the $x$ relic abundance is thus substantially
diluted by the entropy release during the subsequent quark-hadron
phase transition, $x$ decays with a lifetime $\agt 0.1$ s will have an
adverse effect on primordial nucleosynthesis~\cite{lang} (see also
Ref.~\cite{moh}). Thus the interpretation of the KARMEN anomaly as a
singlet neutrino is severely challenged by astrophysical and
cosmological arguments, although there may be a loophole for fast
decays with lifetime $\ll 0.1$~s which occur in the supernova core
and, in the early universe, before nucleosynthesis. Therefore we
proceed to examine the implications of this hypothesis for laboratory
experiments which can test it definitively.

Identifying $x$ as mainly sterile has several distinct repercussions
for the $x$-decay modes, since GIM cancellations no longer suppress
FCNC:

(i) The $x\to\nu\nu\nu$ invisible modes are now appreciable; we obtain
\begin{equation}
\Gamma (x\to\nu\nu\nu )/ \Gamma(x\to e^-e^+\nu_e) =
[|U_{ex}|^2 + |U_{\mu x}|^2 + |U_{\tau x}|^2]/|U_{ex}|^2 ,
\label{eq:nununu}
\end{equation}
assuming that just one singlet neutrino flavour takes part and the
other 3 neutrinos are much lighter than $x$; thus the invisible
branching fraction is never less than $50\%$. But our determination
of $|U_{ex}|^2$ from the KARMEN $B/\tau$ plot is preserved, because
the latter is determined by the visible decays only: $\tau =
\Gamma(x\to {\visible})^{-1}$.  Adding invisible decay modes increases
the total $x$-width and hence the total number of decays in a given
detector, but the visible fraction decreases by the same factor and
the number of visible decays remains unchanged (so long as the
distance to the detector remains much less than the mean decay length,
which is the case here).

(ii) Radiative decays $x\to\nu\gamma$,
going via loop diagrams, also escape GIM suppression for
mainly-singlet $x$ ; we obtain
{}~\cite{deruj,hprs}
\begin{equation}
\Gamma (x\to\nu\gamma )/ \Gamma(x\to\nu\nu\nu) =
\frac {27\alpha}{8\pi} = {1\over 128}\ ,
\label{eq:nugam}
\end{equation}
assuming one singlet and summing 3 light final flavours as
before. Radiative decays are thus between $0.4\%$ and $0.8\%$ of total
decays, but can dominate visible decays as follows:
\begin{equation}
|U_{\mu x}|^2+|U_{\tau x}|^2 > 127\ |U_{ex}|^2, \quad
                     x\to\nu\gamma\;{\rm dominates};
\label{eq:ngdom}
\end{equation}
\begin{equation}
|U_{\mu x}|^2+|U_{\tau x}|^2 < 127\ |U_{ex}|^2, \quad
                     x\to e^-e^+\nu\;{\rm dominates}.
\label{eq:eendom}
\end{equation}

(iii) When $x\to\nu\gamma$ modes start to dominate over $x\to
e^-e^+\nu$ in the visible decays, we can no longer compensate an
increase in $x$-production (increase in $|U_{\mu x}|^2$) by a decrease
in $|U_{ex}|^2$, so we have to add Eq.(\ref{eq:eendom}) to the
solution constraints for a sterile neutrino. This leaves a reasonable
range for sterile Dirac but only a small region for sterile Majorana
solutions in Fig.1.

(iv) A new class of solution is possible, in which $x\to\nu\gamma$
dominates the visible decays, in contrast to the solutions displayed
in Fig.1 where $x\to e^-e^+\nu$ dominates. These solutions have
$B\times\Gamma({x\to \visible})\simeq 3\times 10^{-17}\; \mu {\rm
s}^{-1}$ with $B$ determined by Eq.(\ref{eq:bf}) as before, but
$|U_{ex}|^2$ now contributes negligibly to the decay and we have
$\Gamma({x\to \visible})=1.2\times 10^{-5} (|U_{\mu x}|^2+ |U_{\tau
x}|^2)\; \mu {\sec}^{-1}$ for Dirac $x$ (double this for Majorana
$x$).  Hence these solutions are characterized by
\begin{equation}
|U_{\mu x}|^2 (|U_{\mu x}|^2 + |U_{\tau x}|^2) \simeq 0.8\times 10^{-10}
\; {\Dirac},\quad 0.4\times 10^{-10} \; {\Majorana}\ ,
\label{eq:nusol}
\end{equation}
which automatically satisfies the requirement $|U_{\mu x}|^2 < 2
\times 10^{-3}$ from Eq.(\ref{eq:shrock}). They are also constrained
by the preliminary bound on $\nu_\tau\to\nu_i\gamma$ decays~\cite{wa66}
from the BEBC-WA66 beam dump experiment~\cite{coop}, which here translates
into
\begin{equation}
|U_{\tau x}|^2 (|U_{\mu x}|^2 + |U_{\tau x}|^2) \alt 0.016\; {\Dirac},
\quad 0.008\; {\Majorana}\ .
\label{eq:wa66}
\end{equation}
The $x$ mean lifetime, dominated by $x\to\nu\nu\nu$ decays, can range
between about $50$~s (at the limit where $|U_{\mu x}| >> |U_{\tau
x}|$) to about $5\times 10^{-3}$~s (where $|U_{\tau x}|$ approaches its
upper limit from Eq.(\ref{eq:wa66})). Thus the $x$ mean decay length is
always much greater than the distance to the KARMEN detector, as
required. Also at the lower end of this lifetime range, the cosmological
and astrophysical constraints may perhaps be evaded.

We now address the further implications of the $x$-neutrino scenario
for laboratory experiments. This scenario implies that, with the
stopped $\pi^+$ decay chain as the neutrino source, $x$ will be
produced not only via $\pi^+\to\mu^+x$ (giving the time-signatured
anomaly in the KARMEN detector), but also via $\pi\to e^+x$ and
$\mu^+\to \bar\nu_\mu x e^+, \bar x \nu_e e^+$ channels. It is
interesting to ask how many of these $x$ should decay in the KARMEN
detector (or other detectors), compared to the anomaly events, and
what their signatures may be.

(a) The $\pi^+\to e^+x$ channel.  The rate depends on $|U_{ex}|^2$.
The fraction that decay in a given detector depends inversely on
the momentum with which $x$ is produced. Hence the ratio of
detected $x$ decays from this channel compared to
``anomaly" events from $\pi^+\to\mu^+x$ is
\begin{equation}
\frac {N(\pi\to ex:\; x \; {\detected})}{N(\pi\to\mu x:\; x \; {\detected})}
= \frac{|U_{e x}|^2[m_\pi^2(m_e^2  +m_x^2)-(m_e^2  -m_x^2)^2]}
     {|U_{\mu x}|^2[m_\pi^2(m_\mu^2+m_x^2)-(m_\mu^2-m_x^2)^2]}
\simeq 0.15\ \frac{|U_{ex}|^2}{|U_{\mu x}|^2}
\end{equation}
In these events $x$ has velocity $\beta_x=0.89$ and reaches the
detector essentially in coincidence with the prompt $\nu_\mu$ burst
from the initial pion pulse, well within the resolution defined by the
100 ns pulse length at the KARMEN source.  For solutions with dominant
$x\to e^-e^+\nu$ visible decays, the spectrum of visible energy (summed
$e^+e^-$ kinetic energies) is shown in Fig.2, including initial $x$
polarization and folding in the relative decay probability in a
detector.  About $76\%$ of such detected decays have visible energy
greater than 40 MeV, whereas each prompt $\nu_\mu$ carries only 29.8
MeV total energy, so these $x$-decay events should be quite
distinctive.  For the more restricted Majorana solutions,
$|U_{ex}|^2/|U_{\mu x}|^2 \alt 10^{-2}$ and the $\pi\to ex$ signal is
strongly suppressed.  Solutions with dominant $x\to\nu\gamma$ visible
decays necessarily have $|U_{ex}|^2 << |U_{\mu x}|^2 + |U_{\tau
x}|^2$; if this is achieved with $|U_{\mu x}|^2 < |U_{ex}|^2 <<
|U_{\tau x}|^2$, then the $\pi\to ex$ production channel will be
important, giving decay photon energies from 4 to 70 MeV with mean
energy 26 MeV, to be compared with a narrow spike at $E_{\gamma}=17$
MeV from anomaly events.

(b) The $\mu^+\to \bar\nu_\mu x e^+$ channel.  Once again the rate depends
on $|U_{ex}|^2$ and the fraction decaying in a given detector depends
inversely on the $x$-momentum.  Numerical calculations give the
detected event ratio:
\begin{equation}
\frac {N(\pi^+\to\mu^+\to\bar\nu_\mu e^+ x:\; x \; {\detected})}
      {N(\pi\to\mu x:\; x \; {\detected})}
\simeq 0.38\ \frac{|U_{ex}|^2}{|U_{\mu x}|^2} .
\end{equation}
There is a spread of velocities $0<\beta_x < 0.95$, but these events
are smeared anyway by the parent muon lifetime and arrive with
essentially the same time distribution as the $\bar\nu_\mu$ and
$\nu_e$ interaction events.  The detected visible energy spectrum
for solutions with dominant $x\to e^-e^+\nu$ , calculated for the
$\mu\to x\to ee\nu$ cascade with full spin-correlated decay matrix
elements~\cite{bop}, is shown in Fig.2; it has mean
value $<T_{\visible}>\; \simeq 26$ MeV. This spectrum roughly
resembles the $^{12}$C$(\nu_e,e^-)X$ spectrum shown in Ref.~\cite{kar},
but has a longer tail above 35 MeV, so it should be possible to
distinguish one from the other. As with channel (a) above, this
channel is suppressed by $|U_{ex}|^2/|U_{\mu x}|^2 \alt 10^{-2}$ for
Majorana $x\to e^-e^+\nu$ solutions, but is important for a subset
of $x\to\nu\gamma$ solutions , for which the decay photon energies
extend from 5 to 53  MeV with mean value 26 MeV.

(c) The $\mu^+\to \bar x \nu_e e^+$
channel. Here the rate depends on $|U_{\mu x}|^2$ instead, like
the ``anomaly" channel $\pi\to\mu x$. Numerical calculations
give the detected event ratio:
\begin{equation}
\frac {N(\pi^+\to\mu^+\to\bar x\nu_e e^+ x:\; \bar x \; {\detected})}
      {N(\pi\to\mu x:\; x \; {\detected})}
\simeq 0.40 \; .
\end{equation}
As with channel (b), these events have essentially the same time
distribution as the $\bar\nu_\mu$ and $\nu_e$ interaction events; on
the other hand, the number of these events compared to time-anomaly
events is now firmly predicted.  The detected visible energy spectrum
for $x\to e^-e^+\nu$ solutions is shown in Fig.2; like case (b) it has
mean value $<T_{\visible}>\; \simeq 26$ MeV and can be distinguished from
conventional $^{12}$C$(\nu_e,e^-)X$ by the tail above 35 MeV.  For
$x\to\nu\gamma$ solutions, the decay photon again has energies
between 5 and 53 MeV with mean value 28 MeV.

To summarize, a mainly-sterile neutrino interpretation for $x$ is
consistent with all laboratory constraints, within limited ranges of
mixing parameters, for both Dirac and Majorana options, although there
are some astrophysical and
cosmological problems. Solutions with dominant $x\to e^-e^+\nu$ visible
decays are constrained by
Eqs.(\ref{eq:shrock})-(\ref{eq:sol}),(\ref{eq:eendom}); here the mixing
parameters $|U_{\mu x}|$ and $|U_{e x}|$ are adjusted to give compatible
values of branching fraction $B$ and visible-mode lifetime $\tau$ in
Fig.1 (while $|U_{\tau x}|$ is negligible) .  Alternative
solutions with dominant $x\to\nu\gamma$ visible decays are
constrained by
Eqs.(\ref{eq:shrock}),(\ref{eq:ngdom}),(\ref{eq:nusol})-(\ref{eq:wa66});
here $B$ and $\tau$ are determined by $|U_{\mu x}|$
and $|U_{\tau x}|$  instead (while $|U_{ex}|$ is negligible).
Such interpretations imply that other
sources of $x$-production should contribute appreciable additional
$x$-decay events in the KARMEN detector, with different
time-structures; in channels (a) and (b) the event rate depends on
$|U_{ex}|$ and is appreciable in some Dirac $x\to e^-e^+\nu$ solutions
and some $x\to\nu\gamma$ solutions, but in channel (c) the event rate
is always $40\%$ of the anomaly event rate.  Similar $x$-decay signals
should also appear in the LSND detector~\cite{lsnd}.  Since both
detectors have approximately the same density, and since both
$\nu$-interaction and $x$-decay events have the same inverse-square
dependence on distance $L$ (so long as $L << \beta_x \gamma_x \tau_x
c$), the ratio of interactions to decays should be approximately the
same in both experiments. Such $x$ decays could conceivably provide a
background to the $\bar\nu_\mu\to\bar\nu_e$ oscillation search
currently under way~\cite{osc}, although the $\bar\nu_ep\to
e^+n$ signal is distinguishable in principle by detecting the delayed 2.2
MeV gamma from the subsequent neutron capture $n (p, d)\gamma$. In
high-energy neutrino beams, however, where the parent pions have been
boosted to energies $E_\pi = \gamma m_\pi$ with $\gamma >> 1$, the
number of $x$-decays in a given detector volume will scale down as
$p_x^*/[E_x^* \gamma ]$ (where $p^*$ and $E^*$ are $\pi$-restframe
momentum and energy) while the $\nu$ interactions will scale up as
$\gamma$; thus the fraction that are decays decreases as $1/\gamma^2$
and rapidly becomes negligible. Hence $x\to e^+e^-\nu$ decays cannot
contribute significantly to the apparent $e/\mu$ excess in atmospheric
neutrino events~\cite{kam}, where the parent pions and kaons have
energies of ${\cal O}$(GeV). Similarly, there should be a negligible
$x\to e^+e^-\nu$ contribution in accelerator experiments such as
BNL-E776~\cite{bnl} which set an upper limit on $\nu_\mu\to\nu_e$ and
$\bar\nu_\mu\to\bar\nu_e$ oscillations using GeV neutrinos.

Finally, we note that isosinglet (sterile) neutrinos occur naturally
in $SO(10)$ and $E_6$ GUT models, as members of the basic fermion
families~\cite{gelmini}. A very light ($\approx 10^{-2}$ eV) sterile
neutrino has been suggested as a possible~\cite{desh} or even
necessary~\cite{fuller} participant in solar neutrino oscillations,
while one with a mass of ${\cal O}$(keV) is a good candidate for
`warm' dark matter~\cite{dodelson}. Heavy singlet neutrinos could
cause distinctive lepton-number-nonconserving, lepton-flavour-changing
and lepton-universality-breaking effects in a wide range of laboratory
processes~\cite{heavy}.

\begin{flushleft}{\bf Acknowledgments}\end{flushleft}
RJNP thanks J. Kleinfeller, I. Blair and J. Guy for conversations;
VB thanks R. Imlay for a discussion; SS thanks N.~Booth, R.~Mohapatra
and G.~Raffelt for helpful comments.  This research was supported in
part by the U.S.~Department of Energy under Grant No.~DE-FG02-95ER40896
and in part by the University of Wisconsin Research Committee with
funds granted by the Wisconsin Alumni Research Foundation. SS is a
PPARC Advanced Fellow and acknowledges support from the EC Theoretical
Astroparticle Network.

\vspace{1cm}

\section*{Figures}

\begin{enumerate}

\item{Correlation between the $x$ mean lifetime and production branching
ratio $B~= \Gamma~(\pi\to\mu x)/\Gamma~(\pi\to\mu\nu)$, needed to
explain the KARMEN anomaly; the solid curve is taken from
Ref.~\cite{kar} and the dashed curve is its extrapolation.  The
$|U_{\mu x}|^2$ scale is derived from Eq.(\ref{eq:bf}) for $B$.  The
$|U_{ex}|^2$ scale is derived from Eq.(\ref{eq:t}) for the case of
dominant $x\to e^-e^+\nu$ visible decays with Dirac $x$. The most
stringent laboratory constraints on the $|U_{\alpha x}|$ values are
also shown~\cite{pdg,shrock,bern,delee}.  For Majorana $x$ there is an
additional constraint $|U_{ex}|^2~\;~\alt~\;~6\times 10^{-8}$ and the
$|U_{ex}|^2$ scale moves left by a factor of 2.
\label{fig:fig1}}

\item{Visible energy spectra for $x\to ee\nu$ decays from various sources,
weighted by the inverse of the $x$ momentum for the relative decay
probability in a given detector.  The solid curve denotes the stopped
$\pi^+\to\mu^+x$ source; dotted, dashed, and dot-dashed curves denote
the $\pi^+\to e^+x$, $\pi^+\to\mu^+\to\bar\nu_\mu x e^+$ and
$\pi^+\to\mu^+\to\bar{x}\nu_e e^+$ sources, respectively.
\label{fig:fig2}}

\end{enumerate}

\newpage
\begin{center}
{\LARGE Erratum}\\
\end{center}

\vspace{.2in}

In the case that neutrino $x$ is mainly isosinglet (sterile), we
overlooked neutral-current contributions to the decay $x\to\nu
e^+e^-$, as pointed out by J.~Peltoniemi (hep-ph/9606228). The
corrected width for small mixing is
$$
\Gamma (x\to\nu e^+e^-) = 390K \left[(1 + 4x_W + 8x_W^2)|U_{ex}|^2
 + (1 - 4x_W + 8x_W^2)(|U_{\mu x}|^2+|U_{\tau x}|^2)\right]\;{\sec}^{-1},
$$
where $x_W\equiv\sin^2\theta_W=0.23$, $K=1(2)$ for Dirac (Majorana)
$x$ and the $|U_{ex}|^2$ term includes charged-current contributions.
The visible decay width (including $x\to\nu\gamma$) and total width
then become
$$
\begin{array}{lcl}
\Gamma_{\visible} & = & K \left[920 |U_{ex}|^2
                          + 210 |U_{\mu x}|^2
                          + 210 |U_{\tau x}|^2\right]\;{\sec}^{-1},\\
\Gamma_{\total} & = & K \left[2470 |U_{ex}|^2
                          + 1760 |U_{\mu x}|^2
                          + 1760 |U_{\tau x}|^2\right]\;{\sec}^{-1}.
\\
\end{array}
$$
The KARMEN event rate determines the product
$B~(\pi^+\to\mu^+x)\Gamma_{\visible}\simeq 3\times 10^{-11} {\sec}^{-1}$
with $B=0.0285~|U_{\mu x}|^2$ as before, which allows different types
of solution:

(A) If $|U_{ex}|^2$ dominates $\Gamma_{\visible}$, we obtain
$|U_{ex}U_{\mu x}|^2\simeq 1.1 K^{-1} \times 10^{-12}$. For Majorana
$x$, no such solutions are compatible both with the constraint Eq.(9)
and with $|U_{ex}|^2$ dominance of $\Gamma_{\visible}$.  For Dirac $x$
there are acceptable solutions near $|U_{ex}|^2\sim |U_{\mu x}|^2 \sim
10^{-6}$; but as $|U_{ex}|$ decreases, $|U_{\mu x}|$ increases and
they merge into category (B).

(B) If $|U_{\mu x}|^2$ dominates $\Gamma_{vis}$, we obtain a fixed
value $|U_{\mu x}|^2\simeq 2.3 K^{-1/2} \times 10^{-6}$. These
solutions allow ranges of $|U_{ex}|,|U_{\tau x}| < |U_{\mu x}|$, but
when $|U_{\tau x}|$ increases further they merge into category (C).

(C) If $|U_{\tau x}|^2$ dominates $\Gamma_{\visible}$, we obtain
$|U_{\mu x} U_{\tau x}|^2\simeq 5 K^{-1} \times 10^{-12}$ , with $1.5
K^{-1} \times 10^{-6} < |U_{\tau x}|^2 < 1$, where the upper
(unitarity) limit cannot be approached closely because $x$ cannot be
the main component of $\nu_\tau$. Solutions (A) and (B) give mean
lifetimes $\tau_x\sim (150-300)~\sec$, but (C) covers the range
$\tau_x\sim (0.001-150)~\sec$.

The neutral-current contributions now guarantee that $x\to\nu e^+e^-$
is always the dominant visible mode.  Eqs.(16)-(17) are invalidated
and some parameters change, but our qualitative conclusions about the
existence of solutions (especially short-lived type-(C) cases) remain
broadly unchanged.

Concerning additional sources of $x$ at KARMEN, the $\pi^+\to ex$ and
$\mu^+\to \bar\nu_\mu xe^+$ channels depend on the ratio
$|U_{ex}/U_{\mu x}|^2$ as before; this ratio is $\alt 1$ for (A) and
(B) but can be large for type-(C) solutions.  The $\mu^+\to \bar
x\nu_e e^+$ rate is fixed and unchanged.

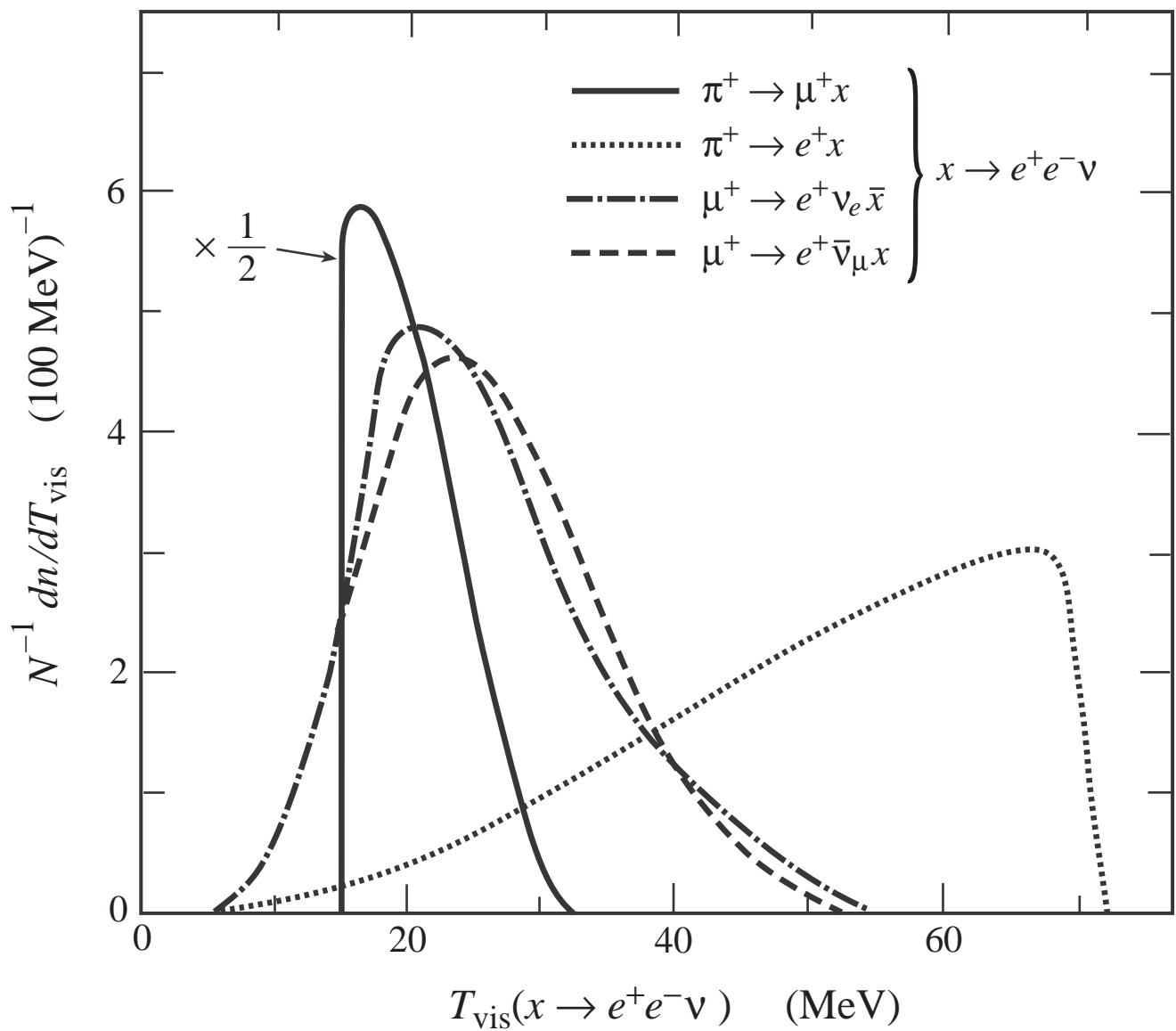
\end{document}